\documentclass[amsmath,amssymb,aps,nofootinbib,onecolumn, notitlepage]{revtex4-1}

\usepackage{graphicx}
\usepackage{dcolumn}
\usepackage{bm}
\usepackage{siunitx}
\usepackage{color}
\usepackage{afterpage}
\DeclareSIUnit\micron{\micro\metre}
\DeclareSIUnit\microliter{\micro\liter}

\begin{document}

\preprint{APS/123-QED}

\title{Particle monolayer assembly in evaporating salty colloidal droplets}

\author{Myrthe A. Bruning}
\email{m.a.bruning@utwente.nl}
\author{Laura Loeffen}
\author{Alvaro Marin}
 \email{a.marin@utwente.nl}

\affiliation{%
 Physics of Fluids Group, Faculty of Science and Technology, Mesa+ Institute, University of Twente, 7500 AE Enschede, The Netherlands
}%

\newcommand{\SImum}{\textrm{\textmu{}m}}

\date{\today}

\begin{abstract}

Ring-shaped deposits can be often found after a droplet evaporates on a substrate. If the fluid in the droplet is a pure liquid and its contact line remains pinned during the process, the mechanism behind such ring-shaped deposition is the well-known coffee-stain effect. However, adding small amounts of salt to such a droplet can change the internal flow dramatically and {consequently} change the deposition mechanism. Due to an increase of surface tension in the contact line region, a Marangoni flow arises which is directed from the apex of the droplet towards the contact line. As a result, particles arrive at the contact line following the liquid-air interface of the droplet. Interestingly, the deposit is also ring-shaped, as in the classical coffee-stain effect, but with a radically different morphology: particles form a monolayer along the liquid-air interface of the droplet, instead of a compact three-dimensional deposit.
Using confocal microscopy, we study particle-per-particle how the assembly of the colloidal monolayer occurs during the evaporation of droplets for different initial concentration of sodium chloride and initial particle dilution. Our results are compared with classical diffusion-limited deposition models and open up an interesting scenario of deposits via interfacial particle assembly, which can easily yield homogeneous depositions by manipulating the initial salt and particle concentration in the droplet.

\end{abstract}

\maketitle


\section{Introduction}

A sessile droplet containing a dilute solution of particles slowly evaporates while keeping its base radius constant. Nowadays we have tools not only to predict where the particles will end up \cite{Deegan1997}, we can also predict the morphology and packing of the particles \cite{Marin2011} (as long as the particle size is uniform enough). Under these conditions, a capillary flow is generated towards the droplet's rim to replenish liquid and maintain the contact line immobile. Consequently, particles follow streamlines that end at the contact line \cite{Deegan1997}, forming the well-known ring-shaped stain (or coffee-stain). The ordering of the resulting three-dimensional arrangement of particles in this ring depends on the particle size and on the flow velocity, which diverges in the last stages of the droplet's life \cite{Marin2011}. 

Nonetheless, small variations in the liquid composition can {drastically change} the system behaviour and also its predictability \cite{Still2012}. From a fluid-dynamical perspective, the main effect of any compositional change is to induce stresses at the liquid-air interface that can influence critically the otherwise dominating coffee-stain effect. Such interfacial stresses can be also induced by thermal gradients, which may occur naturally during the evaporation process \cite{Hu:2005Marangoni,Hu:2006Marangoni}, or may be externally induced \cite{Kita:2016ke,Rossi2019interfacialflow}. Such interfacial stresses induce an internal flow that might eventually overcome the canonical capillary flow. When this scenario occurs, \emph{i.e.} when the coffee-stain effect is not the dominant flow, most of the streamlines close themselves and the destiny of a random particle depends on several factors: its size, shape and density, the distance to neighboring particles, its vicinity to the liquid surface or to the solid substrate, its chemical affinity to all those, etc. In recent years, a vast literature has bloomed with studies dedicated to study the effect of different combinations of such parameters, with special interest in suppressing the ring-shaped stain \cite{Yunker2011,Trantum:2014,Kim2016,Malinowski:2018}.

\begin{figure*}
  \includegraphics[width=\textwidth]{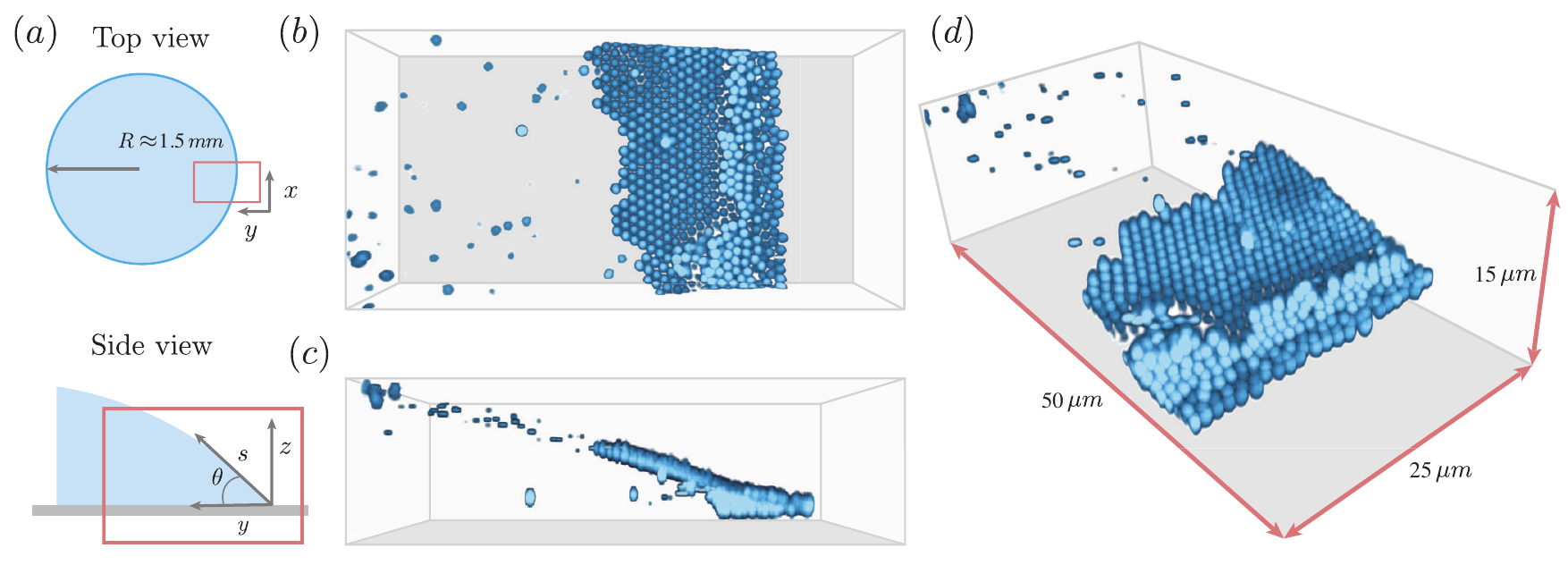}
\vspace{-3mm} \caption{Laser scanning confocal microscopy is employed to obtain three-dimensional information on the particle assembly. Figure (a) shows sketches of the top-view (glass-slide plane) and side-view projection of the scanned volume within the droplet, indicating length scales and axis definitions. The pink box shows the approximate location of the scanned volume respect to the droplet (not in scale).
Figures (b) and (c) show projections of the scanned volume in the $(x,y)$-plane and the $(y,z)$-plane respectively. Figure (d) shows an orthogonal projection of the full scanned volume as obtained from the fluorescent channel, for $C_0$ = 10 mM and ${t/t_{evap}}$ = 0.8. The dimensions of the scanned volume are included.  \label{fig: confocal_overview} }

\end{figure*}
Interestingly, certain type of compositional changes also result in ring-shaped stains despite a substantial change in the internal flow profile: the addition of tiny amounts of NaCl to a water droplet induces an internal flow that looks as an inverted \emph{coffee-stain} flow but also leads to a ring-shaped stain \cite{Marin2019}. The effect is due to a solutal Marangoni flow caused by the presence of NaCl which increases the surface tension of water (as other similar chaotropic electrolytes, it effectively acts as an \textit{anti-surfactant} \cite{Levin2001, Onsager1934}). When a droplet containing a small amount of salt evaporates, its contact line region enriches in salt content over time \cite{Soulie2015}. This induces a surface tension gradient along the droplet of the surface, and a Marangoni flow sets up directed from the apex of the droplet towards the contact line. In a recent paper, Marin \textit{et al.}  \cite{Marin2019} showed both experimentally and numerically, that this surface flow is strong enough to overcome the bulk capillary flow. They also showed that the ring-shaped deposit originates from particles adsorbed at the liquid-air interface (from now on, droplet's surface), and convected towards the contact line region by the surface flow. This evaporation-driven particle assembly mechanism has not been studied in detail, and it will be the focus of the current work.

In this study, making use of laser scanning confocal microscopy, we analyze the particle assembly in colloidal droplets containing small amounts of sodium chloride. Scanning a small volume of the contact line region with high spatial and temporal resolution we confirm that the assembly occurs in a monolayer parallel to the droplet's surface. Confocal microscopy allows us to reconstruct particle-per-particle the formation and growth of the particle monolayer, and investigate the influence of the initial salt molarity and particle concentration.


\section{Experiments}

In a typical experiment, a sessile droplet of about  3 \si{\uL} containing certain amount of sodium chloride and polystyrene colloidal particles is evaporated and simultaneously scanned with a laser scanning confocal microscope. The initial concentrations of sodium chloride ($\geqslant$99.5$\%$ purity, BioXtra) employed are $ C_0=$ 5, 10 and 50 mM (molar mass NaCl is 58.443 g/mol). This choice is made to be as far as possible from the saturation concentration in water (roughly 6 M) and high enough to be able to have sufficient particle accumulation during the time of a typical experiment (about 25 minutes). Experiments for each salt concentration are repeated at least 5 times.  

The colloids employed are polystyrene particles, functionalized with sulfate groups to convey stability to the suspension (microParticles GmbH, PS-FluoGreen-1.0) and died with certain green-emitting fluorescent dye (ex/em: 502 nm/518 nm). The particle diameter is chosen around 1 \si{\um} (namely 1.08$\pm${0.04~}\si{\um}), as a compromise to reduce sedimentation during the fairly long evaporation process (polystyrene density is 1.05 g/cm$\mathrm{^3}$, which makes them almost neutrally buoyant) and large enough to be visible under a confocal microscope for the magnifications employed. {The polystyrene particle concentration in the droplets is 0.01$\%$w.} The suspensions are prepared between 0.5-6 hours before use. Prior to the experiment the solution is sonicated for 60 seconds and left to reach room temperature for 4 minutes. A droplet from this solution of 3 \si{\uL} is then gently pipetted on a Menzel-Gl\"azer circular glass microscope cover slide (\#1.5), previously rinsed with Milli-Q water, dried with nitrogen gas and heated at 80$^{\circ}$C for 2.5 minutes. This treatment ensures pinning of the droplet throughout its evaporation time. The initial contact angle of a salt water drop on this substrate is 70$^{\circ}$.

\begin{figure*}
  \includegraphics[width=1\textwidth]{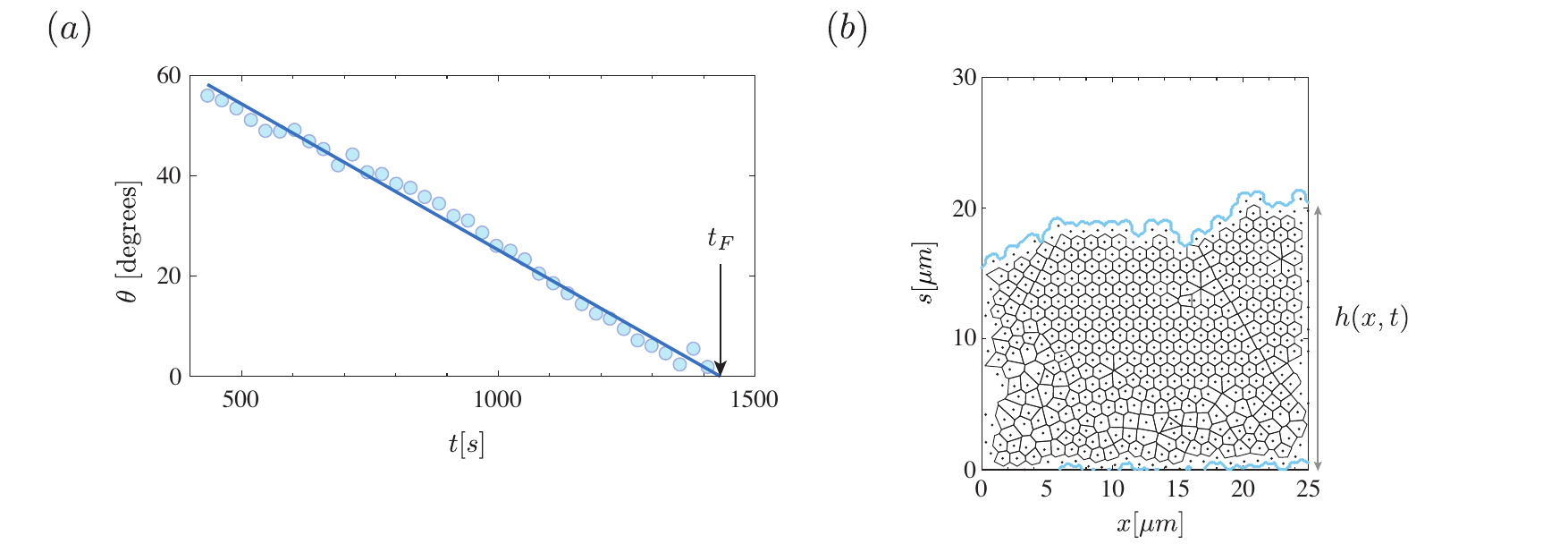}
\vspace{-3mm} \caption{Overview of the typical analysis results. Figure (a) shows the contact angle $\theta$ of the droplet as a function of time. A linear fit through this data is used to access the final evaporation time $t_F$, defined at $\theta$ = 0$^{\circ}$. Figure (b) shows an overview of the analysis results for the same confocal scan as shown in Fig. \ref{fig: confocal_overview} ($C_0$ = 10 mM NaCl and $\tilde{t}$ = 0.8). The found particle positions are shown as as black dots. These particles are plotted along $s$, which is the axis along the droplet's interface, for which the droplet contact angle $\theta$ is used. The contour of the droplet is given as the outer light blue line, and the typical detected Voronoi cells are presented. The definition of the height of the contour $h(x,t)$ is given as well.}  \label{fig: Analysis} 

\end{figure*}

The evaporation experiment is performed in a controlled environment with T = 23$\pm$1\si{\celsius} and RH = 50$\pm 1\%$ in a custom-made chamber. The humidity is controlled using a digital humidity sensor (Glyduino AM2302 DHT22), connected to an Arduino micro-controller. Depending on the outside humidity versus the target humidity, either dry or wet air is supplied into the chamber. Nitrogen gas serves as dry air, while humid air is created by bubbling gas through a water column. The flow rate of the incoming air is maintained low to avoid air convection inside the chamber, which would influence the evaporation of the droplet. The total evaporation time of the droplets is 26$\pm$2.5 minutes.

The positions of the colloids are obtained using an inverted laser scanning confocal microscope (Nikon A1R HD) with an oil-immersed objective of 60x magnification, which yields a resolution of \SI{0.10}{\um}/pixel. The total area observed in the object plane is \SI{25}{\um} in width ($x$-axis, along the contact line), \SI{50}{\um} in depth ($y$-axis, towards the droplet's center). The volumetric information is obtained by scanning a total of \SI{15}{\um} along the optical axis ($z$-axis, perpendicular to the glass slide), with steps of \SI{0.35}{\um} between scanned planes, at a speed of 0.64 s per position. The humidity is manipulated to make sure that the evaporation process is slow enough to prevent blurred images during the time of a full scan (about 30 s). Also note that the step size is chosen to be smaller than the particle diameter, such that each particle appears in multiple layers of the scan for a better resolution in the particle positioning. A blue laser (488 nm) is used to excite the fluorescent particles. Both the fluorescent emission channel and the transmission channel are recorded and used for obtaining the particle positions. Three projections of the scanned volume obtained from the fluorescent channel are shown in Fig. \ref{fig: confocal_overview}, in which the relevant dimensions of the scanned volume and the axis definitions are included. The different projections in the figure clearly show that the particles arrange themselves into a monolayer that extends along the droplet's surface. A more detailed observation to the top-view and side-view projection (see Fig. \ref{fig: confocal_overview}b and c) shows that the particles in the first rows follow a somewhat irregular arrangement, not entirely following the surface. This could be due to some minor contact line motion or due to the finite roughness from the glass substrate. Nevertheless, the well-defined order of the particles and their arrangement along the droplet surface is notorious in the subsequent particle rows. 

\begin{figure*}
  \includegraphics[width=1\textwidth]{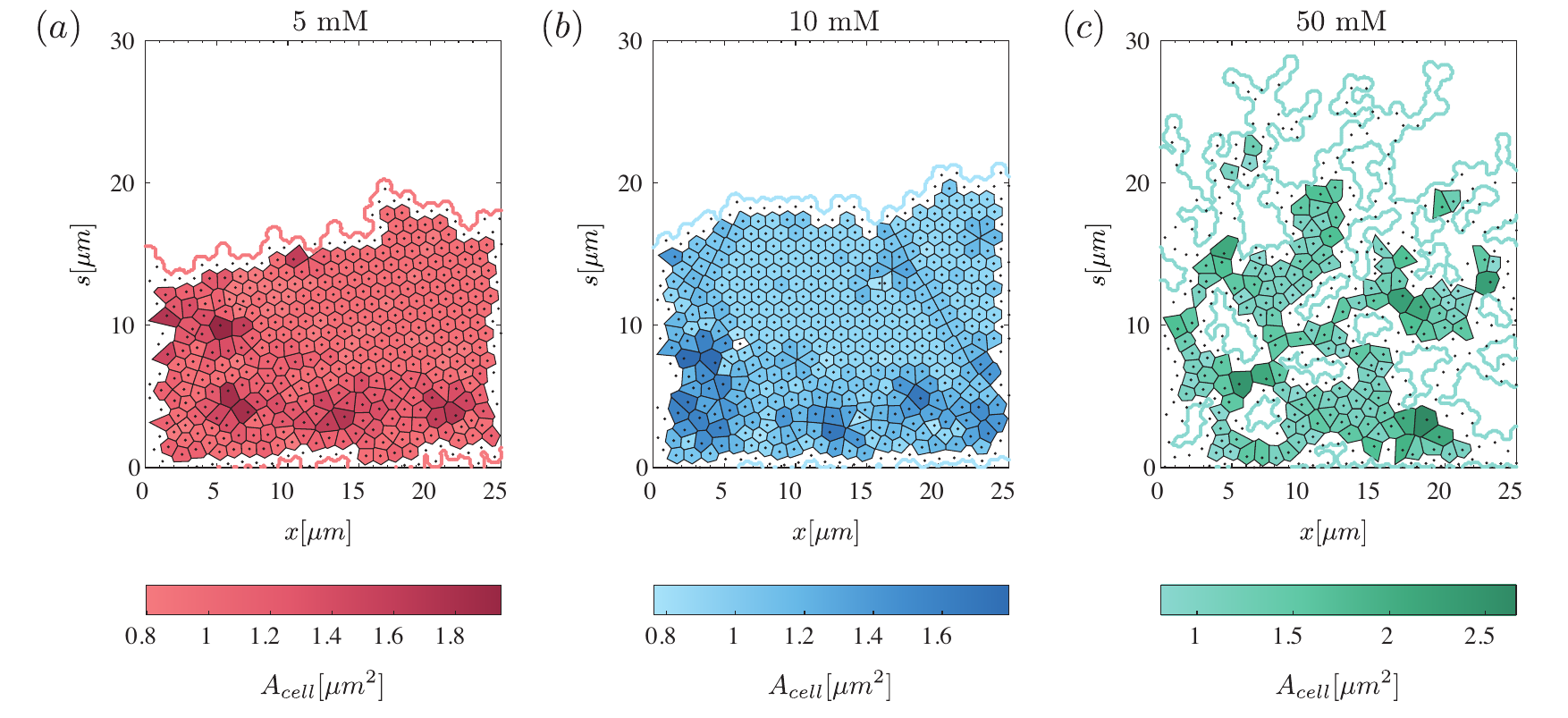}
\vspace{-3mm} \caption{Typical results of a Voronoi cell analysis performed in the particle monolayer for different salt concentrations: (a) 5 mM, (b) 10 mM and (c) 50 mM. The colorbar represents the area of the cell $A_{cell}$ in \SI{}{\micro\meter}$^2$.  \label{fig: Voronoi} }
\end{figure*}

The data obtained from the confocal microscope permit to assess both the order and the growth rate of the monolayer. The former will be done using a 2D Voronoi analysis in the plane of the droplet's surface, the latter by studying the evolution of the upper \emph{contour} of the monolayer that grows towards the center of the droplet. 
First, the 2D Voronoi analysis requires the three-dimensional particle positions. For this we use the output from the transmission channel, as this contains the largest contrast. Using the points with maximum intensity within each layer of the scan, we determine the $(x,y)$-position of each particle. This is done for all layers in every full scan. As each particle is found in multiple layers, the average of these positions is used to find the final $(x,y,z)$-position of this particle. At each time step the contact angle $\theta$ of the sessile droplet is obtained by fitting a plane through the detected particles, which yields a nearly linear decrease of $\theta$ versus time, as can be seen in Fig. \ref{fig: Analysis}a. A linear extrapolation is used to determine the final evaporation time $t_F$, defined at the point where $\theta(t=t_F)$ = 0$^{\circ}$, enabling us to compare different experiments by normalizing time as $\tilde{t} = t/t_F$. Note that $t=$ 0 is defined as the moment at which the droplet is deposited at the substrate, \emph{i.e.} when the evaporation process starts. The contact angle analysis allows us to define an additional axis which we will identify as the $s-$axis (Fig. \ref{fig: confocal_overview}a), which starts at the contact line and is directed towards the droplet's symmetry axis. As shown in Fig. \ref{fig: confocal_overview}a, the origin of the $s$ coordinate is the contact line of the droplet. 

In Fig. \ref{fig: Analysis}b we show an example of the obtained particle positions (black dots) in the $(x,s)$-plane and their corresponding Voronoi cells for $C_0$ = 10 mM and $\tilde{t}$ = 0.8. By definition, all points contained in a particle's Voronoi cell are closer to that particle than to any other. The regularity in the Voronoi cell's area is a quantitative sign of order in the particle array. Figure \ref{fig: Analysis}b also includes the traced upper contour of the particle array as a light blue line, obtained from the fluorescent channel of the scan. The contour is obtained from the reconstructed three-dimensional particle array and projected to the $(x,s)$-plane. It is represented quantitatively by the function $h(x,t)$, which can be obtained with subparticle resolution through our measurements: $h(x,t)$ is defined as the maximum height reached by particles in the array at a given position $x$ and instant $t$, and therefore it yields an unique value for each $x$-position.

\section{Results}


\subsection{Voronoi analysis within particle monolayer}

\textcolor{black}{We perform a Voronoi cell analysis to gain insight in the level of ordering within those particles detected in the monolayer. Figure \ref{fig: Voronoi} shows the results of Voronoi analyses performed for experiments with identical initial particle concentration and increasing initial salt concentrations $C_0=$ 5, 10 and 50 mM of NaCl, all captured at the dimensionless time $\tilde{t}=$0.8. Those cells located at the edge of the image are removed, as their size is undefined due to the absence of neighbors. Clearly the cases of 5 and 10 mM look very similar, with the same degree of order, manifested by large regions with hexagonal lattice structure. In the case of 50 mM the particle distribution is rather sparse, and the detected cells show irregular shapes. These observations are quantified using the normalized cell size $A_{cell}/A_{hex}$, where $A_{hex} = \frac{1}{2}\sqrt{3}D^2$ is the size of a perfect hexagonal cell ($D$ is the particle diameter). 
}

\begin{figure*}
  \includegraphics[width=1\textwidth]{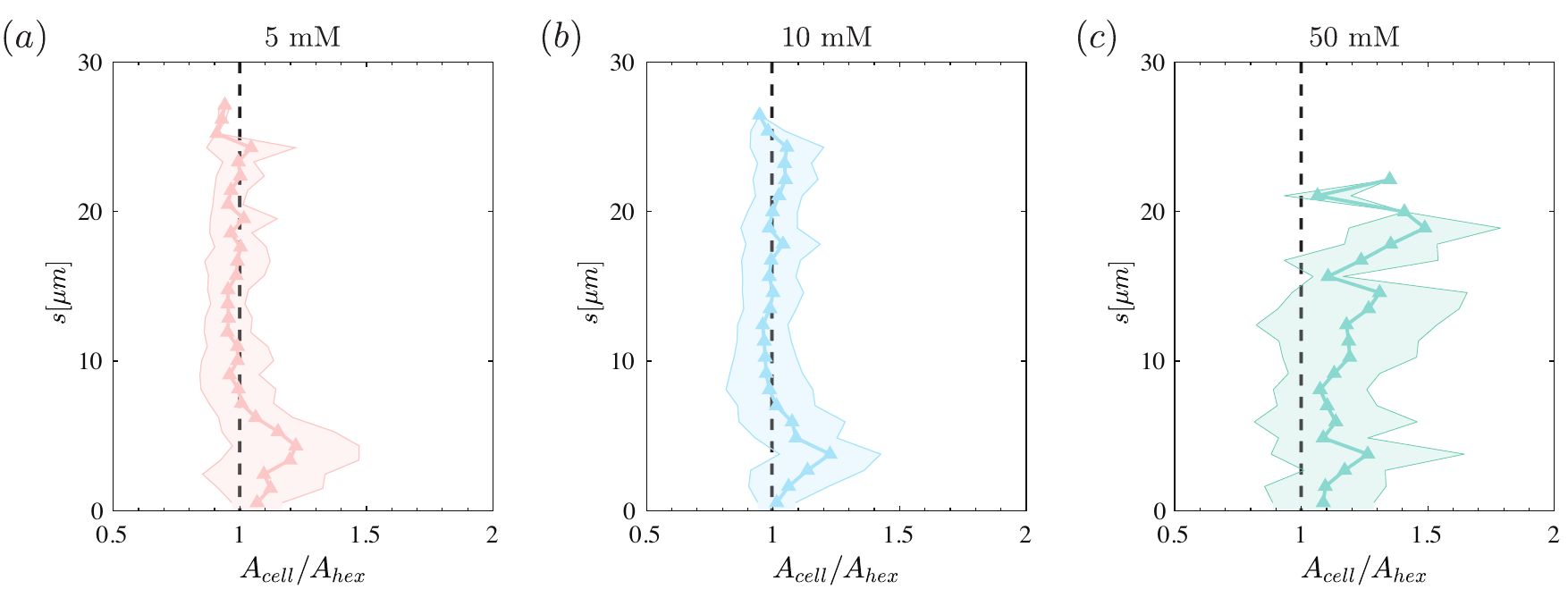}
\vspace{-3mm} \caption{The normalized average size of the Voronoi cell as a function of the distance from the contact line is shown for three different salt concentration, (a) 5 mM, (b) 10 mM and (c) 50 mM. For each concentration, the average cell size of five experiments is used, the shaded area displays the standard deviation. All structures are analyzed at $\tilde{t}$ = 0.8. The cell size is normalized with the area of a hexagonal cell, which is calculated using the particle diameter.  \label{fig: Cell} }

\end{figure*}

Figure \ref{fig: Cell} shows the normalized cell area plotted against the distance from the contact line $s$, for different $C_0$. Each plot includes data from 5 different experiments at identical conditions. The average cell size is measured along $s$ and all data is binned with a fixed bin of size $D$. The shaded region around the curve displays the standard deviation. Two important observations can be made from this figure. First, the particle ordering of the 5 and 10 mM case are almost indistinguishable, and strongly contrast with the 50 mM case. For this highest salt concentration the average cell size is larger as well as the cell size variance. Second, the particle array in the 5 and 10 mM cases exhibit large regions with a hexagonal structure. Near the contact line (\emph{i.e.} for small values of $s$) the normalized average cell size is slightly above 1. As can be seen in Fig. \ref{fig: confocal_overview}c, the particles closest to the contact line are somewhat less ordered and the packing is not entirely two-dimensional. A possible reason might be found on some marginal contact line creeping in the early stages of evaporation that disarranges the particle structure.

Salt-induced aggregation of colloidal particles at air-water interfaces has been studied since decades now \cite{Pieranski1980,Robinson1992:I,Robinson1992:II,Robinson1992:III}, often motivated as a model experiment to study diffusion-limited aggregation \cite{Witten1981:DLA}. Note however that our particles arrange themselves at much shorter distance than in the so-called \emph{colloidal crystalline} structures \cite{Pieranski1980}, which could be easily due to the presence of electrolyte in all our experiments.
Our system shows however more analogies to those scenarios in diffusion-controlled deposition models \cite{Meakin1983:Diffusion,Racz1983:Deposition}, specially due to the two-dimensionality of the particle monolayer. Measurements in that direction will be shown in the following section, in which we asses the growth of the monolayer as a deposition process.


\subsection{Particle monolayer growth and contour roughness}

\begin{figure*}
  \includegraphics[width=1\textwidth]{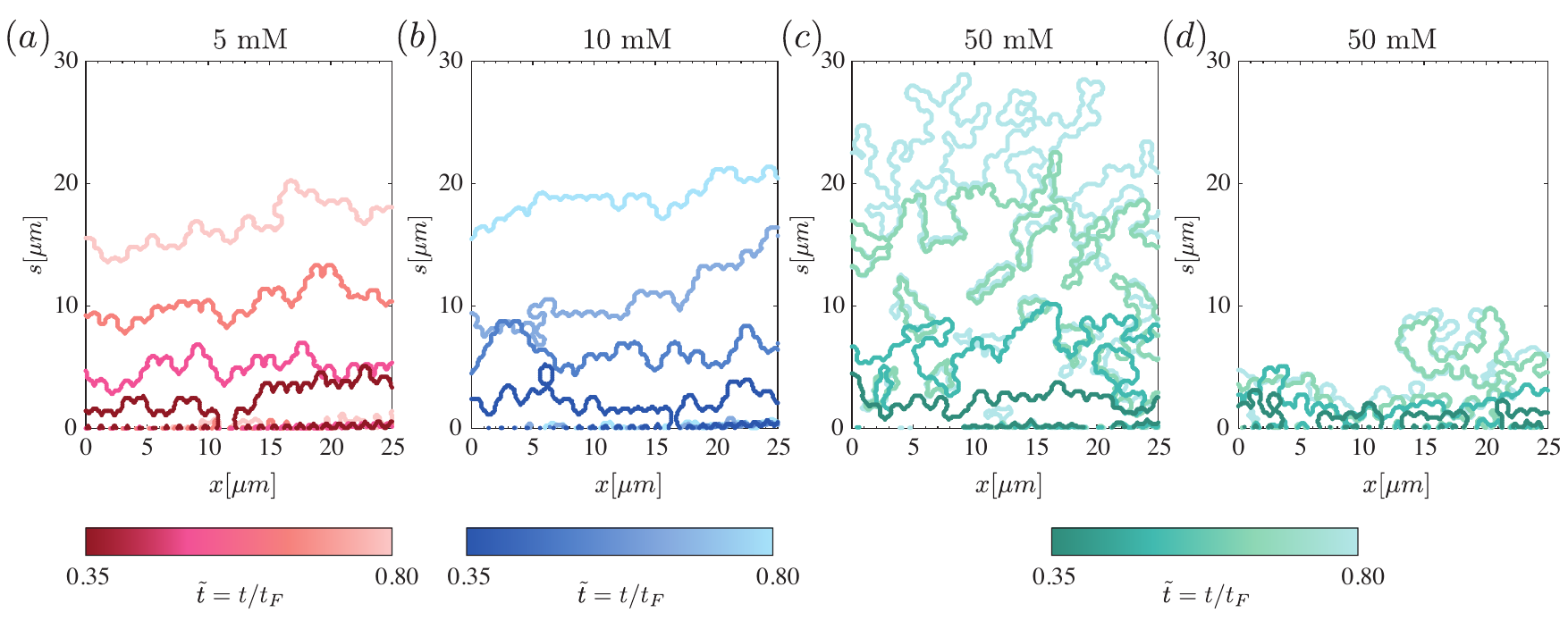}
\vspace{-3mm} \caption{Contour height $h(x,t)$ of particle monolayers along the droplet surface as a function of dimensionless time $\tilde{t}$,  for three different initial salt concentrations.  Figure (a) shows 5 mM, (b) 10 mM and both (c) and (d) show examples of 50 mM, as large heterogenities are observed for different positions along the droplet's contact line. For each experiment contours are shown for $\tilde{t}$ = 0.35,  0.5, 0.65, and 0.8.\label{fig: Contours} }
\end{figure*}

We characterize the growth of the monolayer by studying the evolution of its contour height $h(x,t)$, defined as the maximum distance of the particle structure from the contact line for each position $x$ at every instant $t$. Fig. \ref{fig: Contours} shows the evolution of the contours of the particle monolayer for different initial salt concentrations. The colorbar represents the dimensionless time of the experiment.  We show contours up to $\tilde{t} = 0.8$, since final stages of evaporation yield less reliable results to compare due to depinning events. The cases of $C_0$ = 5 and 10 mM yield almost identical contours, growing fairly uniformly. For the 50 mM case however, we observe very different behavior in different positions along the contact line. Regions depleted of particles as shown in Fig. \ref{fig: Contours}d coexist with highly populated regions as in Fig. \ref{fig: Contours}c. The contours in both Fig \ref{fig: Contours}c and d, but mainly the former,  present prominent finger-like structures and boundant holes that are completely absent in experiments with 5 and 10 mM NaCl.

Another important observation is the way in which the particles attach to the monolayer. In the case of low salt concentration (5 and 10 mM) all particles arrive one by one, as single particles, whereas for the high salt concentration (50 mM) a significant amount of the particles are already agglomerated into small clusters before attaching to the monolayer. The increase of electrolyte in solution screens the electrostatic barrier that prevents the particles from aggregation \cite{Prieve1986}. This is consistent with classical studies on salt-induced aggregration \cite{Pieranski1980,Robinson1992:I,Robinson1992:II,Robinson1992:III}, and also explains the increase of the Voronoi cell area for 50 mM. The less packed and more intricate particle arrangement for the $C_0$ = 50 mM results in a larger area coverage (\emph{i.e.} larger Voronoi cell area) for the same number of particles when compared with the area coverage for $C_0$ = 5 or 10 mM.


For studying the growth of the monolayer, we average along the contact line axis $x$ to obtain $\bar{h}(t)$. Interestingly, for the cases of $C_0=$ 5 and 10 mM, the particle monolayer grows following roughly a scaling $\bar{h}\sim t^{2.5}$, the same scaling found for the particle count at intermediate stages (discussed further below). While the average contour evolution $\bar{h}$ for 50 mM is much noisier and it is difficult to draw any conclusions on its scaling. Data regarding $\bar{h}(t)$ and extended data on the particle count are shown in Appendix A.
The growth of complex interfaces was an intense topic of research after the celebrated diffusion-limited aggregation model introduced in the early 80s by Witten and Sander \cite{Witten1981:DLA}. In this model, particles aggregate by diffusing to a cluster and adhering to it irreversibly. The same principle for studying pattern formation was soon applied to deposition problems \cite{Meakin1983:Diffusion,Family1985,Kardar1986}: particles arrive randomly to an interface where they will adhere irreversibly following certain rules. The growth of such interfaces was studied intensely but special interest was devoted to study the \emph{roughness} of the formed interface for its potential reactive properties; also referred in the literature as ``surface width'' or ``width of the active zone'' \cite{Family1985}. This roughness $w(t)$ is defined as the standard deviation of the contour height $h(x,t)$ at every instant $t$ (note that we follow the nomenclature used by Family \cite{Family1990}). As proposed by Family \& Vicsek \cite{Vicsek1984DLAscaling,Family1985}, the roughness $w(t)$ follows a dynamic scaling with $w \sim t^{\beta}$ for the initial growth, and saturating to a constant value at large times. Several deposition models were found to follow such scaling, but we will focus here in only two of them: In the simplest of them, particles appear at random $x$ positions and they pile up on top of the previously deposited particle at that position. Since each particle is independent and does not interact with neighboring particles, this model is often referred as \emph{Poissonian}. This simple growth model yields a roughness growth exponent $\beta=$ 0.5, independent of the number of dimensions.

Another way to model the interface growth is through a so-called \emph{Ballistic} process: particles fall down onto the substrate and stick either on the top of the column they initially fell in, or stick to a particle in the nearest neighbor columns. This reflects in a different value for the roughness growth exponent, approximately $\beta = $1/3 (in the 2D-case) \citep{Family1985,Family1990}, which can be directly obtained through the renormalization of the Kardar–Parisi–Zhang (KPZ) equation \cite{Kardar1986}. The idea that our system might follow some of these simplified growth models is very tempting. In a related experimental system, the interfacial clustering of non-spherical colloidal particles observed by Yunker \textit{et al.} \cite{Yunker2013}, the roughness has been proposed to follow universality classes that appears consistent with either KPZ \cite{Kardar1986} or KPZ with quenched disorder (KPZQ) \cite{Amaral1994KPZQ} depending on the particle shape. The experimental evidence has not been entirely conclusive so far \cite{Yunker2013b}, but recent numerical results with patchy colloidal disks \cite{Dias2018} suggests that a transition from KPZ to KPZQ could indeed be possible as the particle aspect ratio increases. 

Looking at the results shown so far in the context of diffusion-limited deposition models, an obvious question arises on whether a similar transition of scaling/model can be measured in our colloidal deposition process. One of the main obstacles to make a direct comparison is the different definition of \emph{time} for experiments and in the models. In order to make a fair comparison with the simulated model, we follow the same approach followed by Yunker \textit{et al.} \cite{Yunker2013} and compare the contour roughness $w(t)$ against the average contour height $\bar{h}(t)$, instead of time, since its origin is typically ill-defined in the experiments. We have simulated the previously described deposition models introducing our empirically found incoming particle rate ($N \sim t^{2.5}$, discussed in a section below), to obtain first-hand numerical data to compare with experiments. Despite the time-dependent particle rate, the simulated \emph{Poissonian} and \emph{Ballistic} deposition models yield roughness scalings as predicted by the Family-Vicsek dynamic scaling $w \sim \bar{h}^{n}$: with $n=$ 0.50 for the Poissonian process and $n=$ 0.31 for the Ballistic, the latter also showing saturation at large $\bar{h}$ ({obtained via 100 simulations using the experimentally found particle rate, see Appendix B for more details}).

In Fig. \ref{fig: Roughness} the experimentally obtained values for the roughness of the contour $w$ are given as a function of its mean height $\bar{h}$, which contains 5 experiments per salt concentration. Given the noisy character of the data and relatively low number of experiments per sample (5 repetitions per salt concentration), it will be difficult to draw conclusions on the scaling of the data. Another drawback of the experimental data is the small range of $\bar{h}$ that can be experimentally surveyed keeping high resolution ($2\lesssim \bar{h}\lesssim 20$ \si{\um}).
Nonetheless, we have proceed to obtain exponential fits trying to reduce bias as much as possible: First, the data has been binned using variable logarithmic bin-widths (shown in Fig. \ref{fig: Roughness}). Second, the number of data points per bins has been extracted. Third, a weighted linear regression is performed on $\log w$ vs $\log \bar{h}$ using the fraction of data points per bin as weights. Using this procedure, we obtain the following exponents: $n =0.41\pm 0.03$ for the 5 mM case, $n = 0.47\pm 0.05$ for 10 mM and $n = 0.37\pm 0.04$ for 50 mM. Other alternative data fitting strategies yield only slightly different results from these results shown here, and they can nonetheless be found in Appendix C. Although the results are not conclusive due to the limitations described above, the obtained exponents fall in the expected trend: the lower initial salt concentration cases 5 and 10 mM develop higher roughness growth exponents (poissonian-like), and smaller roughness growth exponent (ballistic-like) are found for the highest initial salt concentration, which shows more dendritic deposition patterns. A completely different experimental strategy to acquire larger datasets should be followed to confirm this trend, which is out of the scope of the current work.


\subsection{Incoming Particle Assembly Rate}
\begin{figure*}
  \includegraphics[width=0.95\textwidth]{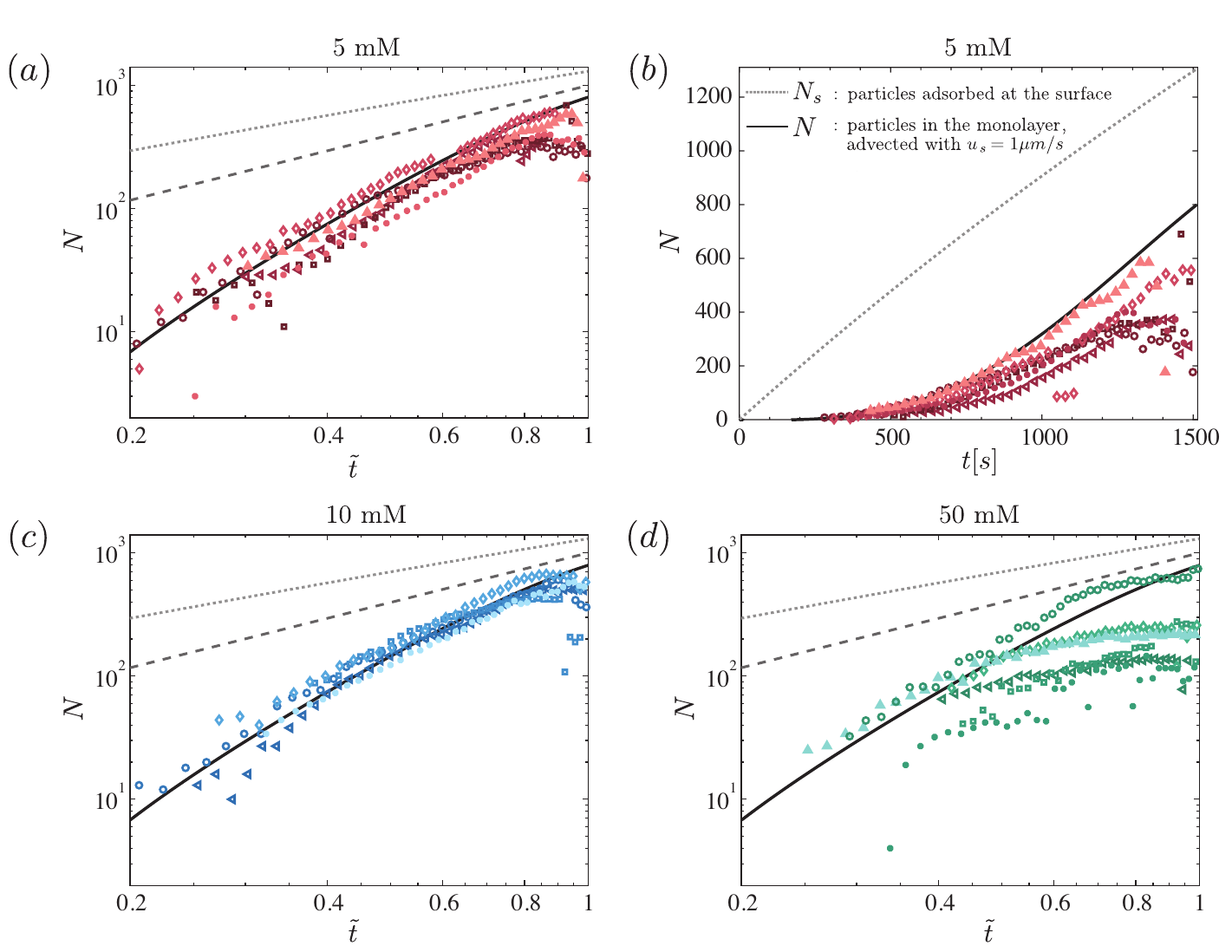}
\vspace{-3mm} \caption{The number of particles $N$ attaching to the monolayer as a function of dimensionless time $\tilde{t}$. Different markers indicate different experiments. Figure (a) shows the results for 5 mM, (c) 10 mM and (d) 50 mM. This number $N$ is the total number of particles which arrive in one measurement section (\SI{25}{\micro\meter} wide). The dashed grey line in these figures represents the classical coffee-stain flow where $N \sim \tilde{t}^{4/3}$ \cite{Deegan1997}, using an arbitrary prefactor. The dotted light gray line in all figures shows the computed total number of particles which are captured by the droplet's surface $N_s(t)$ as a function of time. The results of the numerical solution for $N$ (using a constant surface velocity $u_s =$ \SI{1}{\micro\meter}/s) are given with a solid black line. Figure (b) shows the same curves and experimental data as figure (a) in a linear representation and using real dimensional units. \label{fig:NumberT} }
\end{figure*}

The flux of arriving particles to the monolayer is an important factor to characterize the growth of the deposit. For the classical coffee-stain flow, Deegan {et al.} \cite{Deegan1997} successfully predicted that the total number of particles deposited should scale as $N \sim \tilde{t}^{4/3}$, which is robustly found in experiments whenever this deposition mechanism dominates. In our case of salty droplets, where the deposition is dominated by an interfacial solutal Marangoni flow \cite{Marin2019}, a completely different scaling is expected to be found. Our visualization technique allows to count particle-by-particle with good resolution, and the results are shown in Fig. \ref{fig:NumberT}, where we show the total amount of particles in the monolayer $N$ as a function of dimensionless time $\tilde{t}$. These graphs also include the relation $N \sim \tilde{t}^{4/3}$, the total particle arrival scaling in the classical coffee-stain flow \cite{Deegan1997} (grey dashed lines) as a reference. Clearly the arrival of particles is faster in the case of a salty evaporating droplet. This stresses again the very different deposition mechanism due to the change in flow pattern by the addition of small amounts of salt, and the resulting different particle arrival rate. As with previous results, $C_0$ = 5 and 10 mM show very similar results (Fig. \ref{fig:NumberT}a and c), with very high reproducibility. {The particle number seems to follow approximately a scaling as $N \sim {t}^{2.5}$ (similar to the scaling for $\bar{h}\sim {t}^{2.5}$) for most of the process, but tends to saturate as $\bar{t}\rightarrow 1$, this is shown in Appendix A.}
The case of $C_0$ = 50 mM differs from the previous cases and the measurements also lack the reproducibility shown in $C_0$ = 5 and 10 mM. The lack of reproducibility was discussed previously and it is related to the presence of areas depleted of particles along the contact line, which leads to an early saturation in some of the curves in Fig. \ref{fig:NumberT}d.
In addition, the higher concentration of electrolyte favors the adhesion of particles to the bottom glass substrate due to the screening of the surface charges. Consequently, from the detected particles near the substrate, we estimate that nearly 10\% of the total particles do not succeed to join the particle monolayer in the 50 mM case. This is also shown in Fig. \ref{fig:NumberT}d in which many curves do not reach the same values as those in Fig. \ref{fig:NumberT}a and c.

\begin{figure*}
  \includegraphics[width=1\textwidth]{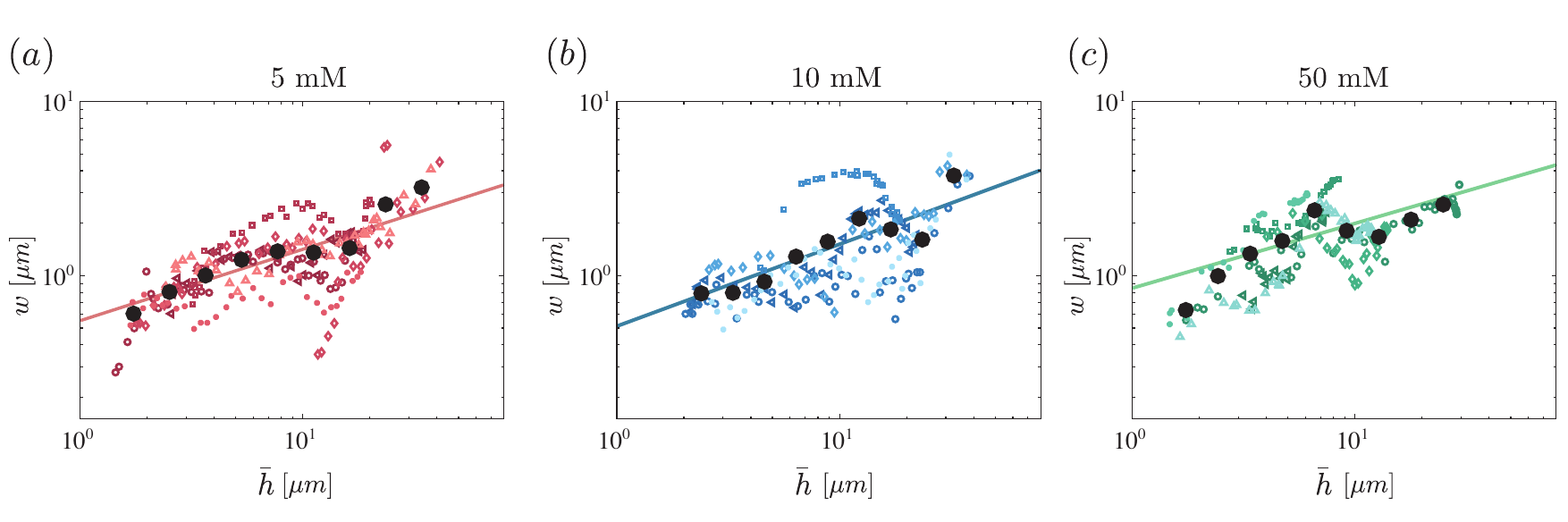}
\vspace{-3mm} \caption{The roughness of the monolayer contours $w$ as a function of the mean height $\bar{h}$. Figure (a) shows 5 mM, (b) 10 mM and (c) 50 mM. Different markers indicate different experiments, the black dots represent the logarithmically binned data. In order to test if the data follows Family-Vicsek dynamic scalings \cite{Family1985} $w\sim \bar{h}^n$, we performed weighted linear regression fits of the binned data. The exponents $n$ found are (a) $n =0.41\pm0.03$, in (b)  $n = 0.47\pm0.05$ and in (c) $n = 0.37\pm0.04$. \label{fig: Roughness}}
\end{figure*}

We propose in the following a simple model to account for the particle number arriving to the monolayer. Since every particle arriving to the monolayer must have been previously adsorbed at the droplet liquid-air surface (or at least in its close vicinity), the number of particles adsorbed at the interface per unit time will give us an upper bound for the number of particles that will eventually end up in the monolayer. 
The particle adsorption mechanism we will consider is the simplest possible: as the droplet's surface retracts keeping a pinned contact line, it will sweep particles along with it. In other words, particles initially dispersed in the bulk will be trapped at the interface as the interface moves downwards \cite{Trantum:2013ffa,Rossi2019interfacialflow,JafariKang:2016}. 
We will not consider any other adsorption mechanism since the particles are functionalized to be practically hydrophilic (which discards any effective diffusivity towards the interface) and their volumetric concentration in the droplet is extremely low. \emph{Diffusiophoresis} is also discarded as a particle migration mechanism in our system since this is essentially a bulk effect that would yield a three-dimensional accumulation of particles in regions of higher salt concentration \cite{Bocquet2008Diffusiophoresis}, and never a two-dimensional monolayer along the surface. Such interfacial particle accumulation has been reported at the same rate for experiments with fresh water \cite{Trantum:2013ffa,Marin2019,Rossi2019interfacialflow}, which is another strong argument against the presence of diffusiophoresis in our system.

As the droplet evaporates, assuming a homogeneous particle distribution in the bulk $C_B$, the decrease in droplet volume $dV/dt$ relates to the total number of particles at the surface $N_s$ as

\begin{equation}
    \label{Eq:dN_dt}
    \frac{dN_s}{dt} = -C_{B}\frac{dV}{dt}.   
\end{equation}

The evaporative flux $dV/dt$ is obtained from the analytical solution of Popov \cite{Popov2005}. The full differential equation for the contact angle $\theta(t)$ is solved for our experimental initial values of volume, contact angle and droplet radius. With the expected $\theta(t)$, the volume loss can easily be calculated at each instant of time. Integrating Eq. \ref{Eq:dN_dt}, we obtain $N_s$, the number of particles swept by the droplet's surface during the evaporation. This gives us an upper bound for the number of particles that might eventually arrive to the monolayer and it is represented in all subfigures in Fig. \ref{fig:NumberT} as a dotted grey line. By definition, this curve overestimates the number of particles at the monolayer.

To compute the arrival rate of particles to the monolayer, we model the particle concentration field along the droplet surface using a simplified 1D transport equation applied to the droplet's surface

\begin{equation}
    \label{Eq:dC_dt}
\frac{\partial C}{\partial t} = J(s,t)-\nabla\cdot (u_s C),
\end{equation}

in which $C(s,t)$ is the particle surface concentration, $J(s,t)$ is the source term for incoming particles and $u_s(s,t)$ is the surface velocity (the advection term). {Using the local volume loss, the local number of particles swept by the interface is known using Eq. \ref{Eq:dN_dt}. }Equation \ref{Eq:dC_dt} is solved numerically using a second order finite difference scheme, details on the numerical scheme and how the spatial variance of $J(s,t)$ is calculated can be found in Appendix D.  Unfortunately, we lack one of the keys to solve this problem: the surface velocity $u_s$. Our own measurements do not have enough time resolution to track particles reliably. In practice, we will assume a constant surface velocity comparable with velocity estimations based on our own experimental data, which are compatible with the experimental data obtained from Marin \textit{et al.} \cite{Marin2019} using 3D particle tracking velocimetry for different salt concentrations. The particle concentration $C(s,t)$ evaluated at the contact line $s=0$, gives us the total number of particles accumulated at the monolayer $N$. 

In Fig. \ref{fig:NumberT}a-d we show the result of the numerical integration of Eq. \ref{Eq:dC_dt} using a constant surface velocity $u_s$ = \SI{1}{\micro\meter}/s as a solid black line (note that we have not performed a systematic study on the optimal value of $u_s$). This solution for $N$ versus $t$ in Fig. \ref{fig:NumberT}b shows two important features: 
First, as expected, particles are being swept by the surface as soon as the evaporation starts, but their arrival to the monolayer is delayed in time. Figure \ref{fig:NumberT}b shows that the first particle is observed experimentally roughly 300 s after the evaporation process begins, which is well captured in the numerical solution for that particular value of $u_s$. 
Second, the curves for the number of particles at the surface $N_s(t)$ (dotted gray curve) and for the number of particles at the monolayer $N(t)$ (dark line) show a very different curvature. This curvature depends strongly on the value of the velocity $u_s$ chosen. The solution for $N(t)$ found with $u_s$ = \SI{1}{\micro\meter}/s matches reasonably well the trend of the experimental data and it is comparable with the values found by Marin \textit{et al.} \cite{Marin2019}. The curvature is related with the time taken for the particle to travel from the point in which it \emph{collides} with the surface to the contact line. Consequently, as we increase $u_s$, the curve for $N(t)$ approaches the curve of particles adsorbed at the surface $N_s(t)$ (for $u_s\gtrsim$\SI{100}{\micro\meter}/s, both curves lay on each other). Despite the limitations of the model, it supports the simple mechanism we propose for the particle assembly at the monolayer: particles are swept by the droplet's surface and consequently advected by the Marangoni flow towards the contact line where they join the monolayer.

\begin{figure*}
 \includegraphics[width=1\textwidth]{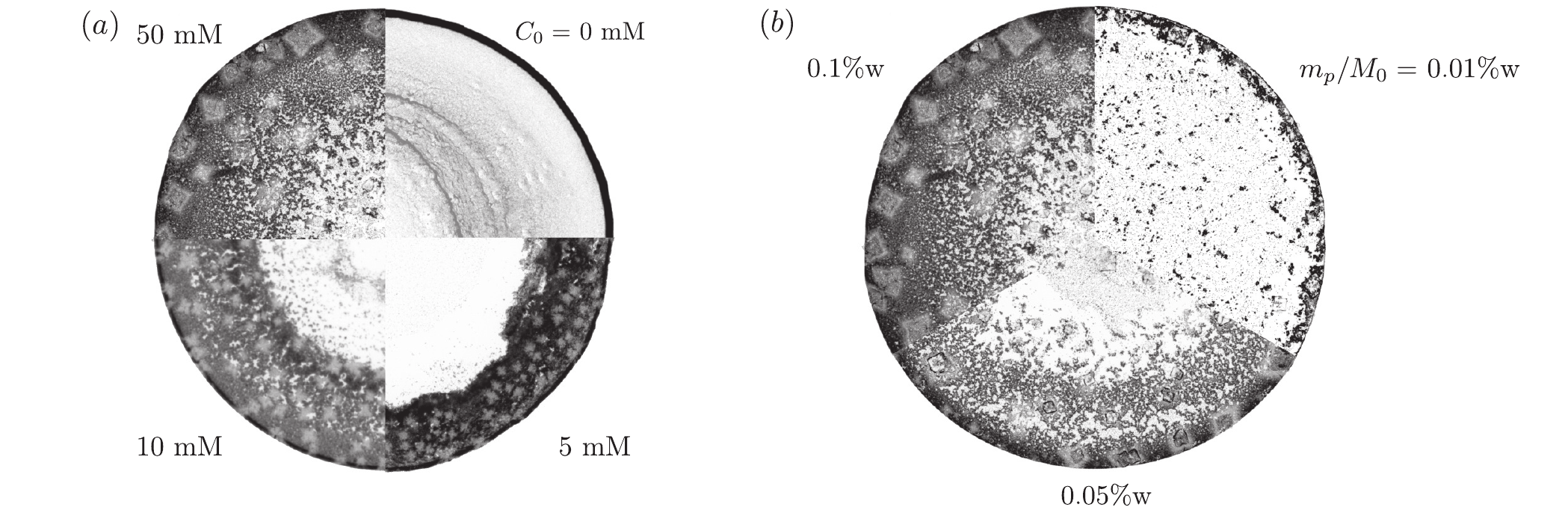}
\caption{ (a) Final stains for different salt concentrations after the liquid phase has completely evaporated. The particle concentration is fixed at 0.1$\%$w/v and only the initial salt concentration $C_0$ increases from 0 to 50 mM. (b) Final stains for different particle concentrations after the liquid phase has completely evaporated. The initial salt concentration $C_0$ is fixed at 50 mM and the particle concentration changes from 0.01 to 0.1$\%$w, defined as the particle mass $m_p$ divided by the initial droplet mass $M_0$.} \label{fig: 2stains} 
\end{figure*}

\subsection{Dried particle deposits}

So far we have shown detailed results on a reduced volumetric region of the droplet, carefully choosing the initial particle and salt concentration to perform the measurements accurately. In this section we analyze the dried deposit of the particles once the evaporation process is completed at a larger scale. In Fig. \ref{fig: 2stains}a we show the resulting stains for increasing salt concentration. In the absence of salt ($C_0=$ 0 mM) the well-known ring-shaped stain appears as a thin black line at the contact line. When the particles are monodisperse enough this ring-shaped stain is characterized by a highly packed deposit, manifested in a thin ring along the contact line. For the same initial number of particles, the presence of 5 mM of salt in the droplet is noticeable by an increase of the ring width. As we increase the salt concentration from 5 to 50 mM, the particle deposition changes drastically and instead of a dark ring of particles, a sparse particle network appears. As described in previous sections, this is formed by a loosely packed monolayer of particles at the droplet's surface, which collapses into the glass substrate when the evaporation process is completed. Interestingly, as the salt concentration is increased, the particle network covers a larger area of the droplet deposit for the same initial number of particles. Salt crystals are clearly observed in the images for this range of salt concentration. Fortunately, the salt crystals only grow in the last stage of evaporation (approximately in the last 3\% of the total time for the highest salt concentration) and is therefore neglected in all our analysis. In Fig. \ref{fig: 2stains}b we have kept the initial salt concentration constant at $C_0=$ 50 mM and increased the initial particle concentration, from 0.01$\%$w/v, yielding an almost complete coverage of the droplet contact area using an initial particle concentration of only 0.1$\%$w/v. Consequently, simply by adding a certain amount of salt and manipulating the initial particle concentration, we can completely suppress the coffee-stain effect.


\section{Conclusions}

In this study we have shown that particles dispersed in an evaporating sessile drop of water with a small amount of salt (NaCl) assemble forming a monolayer that grows from the contact line and along the droplet's surface towards its center. The key ingredient is the solutal Marangoni flow generated by the presence of small amounts of salt \cite{Marin2019}, which generates a surface flow directed towards the contact line. Using laser scanning confocal microscopy we have obtained detailed information of the structure of the particle monolayer. The smallest initial salt concentrations surveyed (5 and 10 mM) yielded well-ordered and highly packed particle arrays. The largest salt concentration (50 mM) resulted in loosely packed and heterogeneous particle monolayers with areas along the contact line depleted of particles.

The roughness of this monolayer (or the \emph{surface width}) has been analyzed using classical diffusion-limited deposition models \cite{Witten1981:DLA,Family1985}. Unfortunately, the lack of enough statistics and the short range of our observation field prevents us to draw conclusions on this matter. Our attempts to fit exponential laws to the roughness have yielded exponents that seem to follow either Poissonian or Ballistic deposition models. But larger data sets would be needed to confirm this trend. The growth of the monolayer is directly proportional to the incoming particle rate, which has been modeled using a simplified surface adsorption and advection equation with constant surface velocity. The model yields incoming particle rates that compare fairly well with that from experiments. This result supports the simple mechanism proposed for particle assembly at the monolayer: particles are swept by the droplet's surface and then advected by the solutal Marangoni flow towards the contact line.

Finally, we have analyzed the dried particle deposits on a larger scale for different initial salt and particle concentrations. As our previous results suggested, at relatively high initial particle and salt concentrations (50 mM), the particle monolayer can easily cover the whole droplet surface and results in an homogeneously and loosely packed particle array when the evaporation process is completed. This mechanism for suppressing ring-shape stains is independent of the particle shape \cite{Yunker2011} and only requires a small amount of salt in the liquid phase.

\section{Acknowledgements} We would like to thank Stefan Karpitschka for several fruitful discussions and comments. Pieter Berghout is acknowledged for his help on the numerical model and Jos\'e Manuel Encarnaci\'on Escobar for his help using the LCSM. MB and AM acknowledge financial support from the European Research Council, ERC-StG-2015 \emph{NanoPacks}, Grant agreement No. 678573.

\section{Appendix}
\subsection{Monolayer growth and incoming particle rate}

\begin{figure*}
  \includegraphics[width=0.85\textwidth,trim={1 15 1 25}]{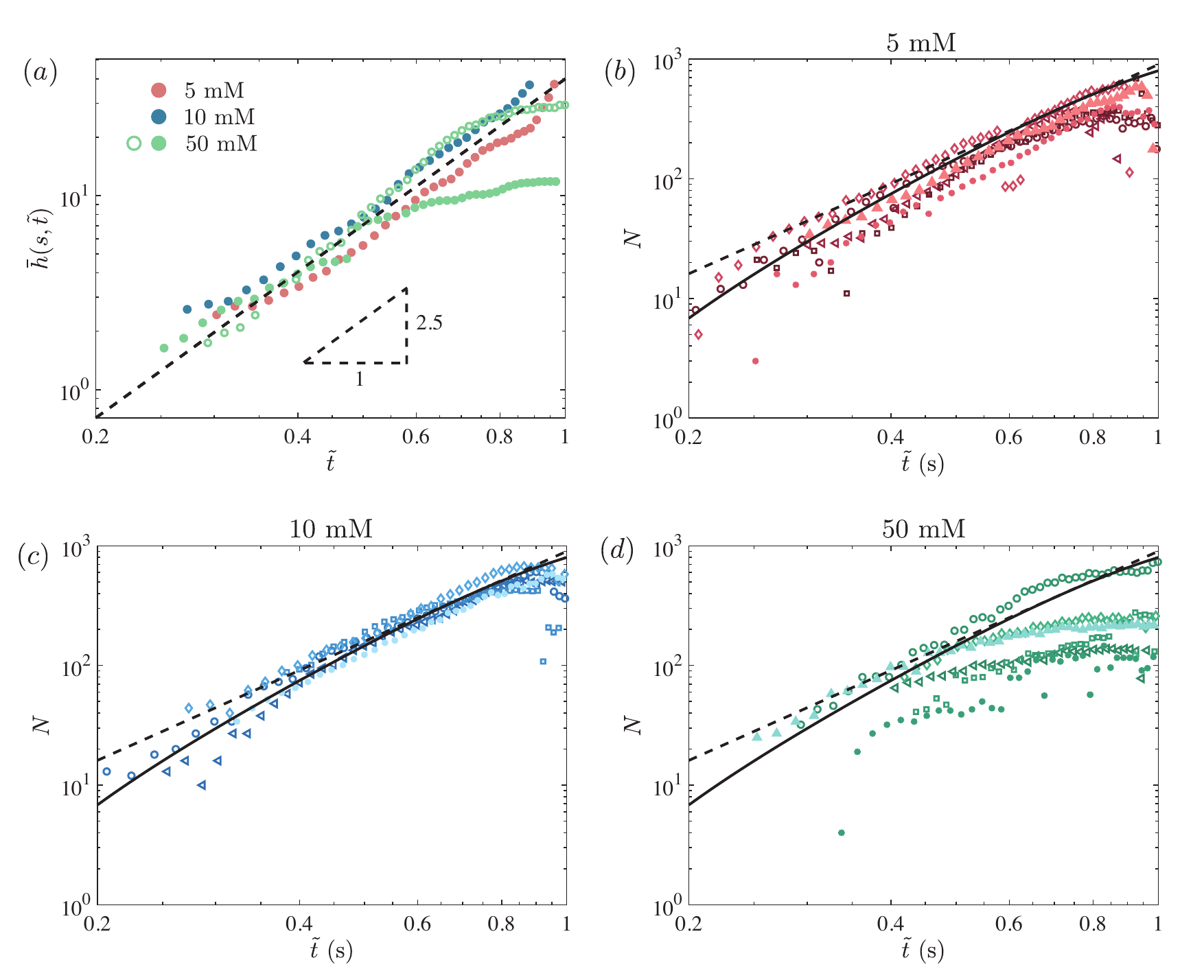}
\caption{Figure (a) shows the typical evolution of the mean height of the monolayer $\bar{h}(t)$ for 4 different experiments with 3 different salt concentrations. Two curves are included for the 50 mM case to highlight the variability found for this salt concentration along the contact line. The dashed black line shows $\bar{h}\sim \tilde{t}^{2.5}$. Figures (b), (c) and (d) show the number of particles $N$ arriving at the contact line for 5 mM, 10 mM and 50 mM respectively. In all 3 figures the solid black curve represents the outcome of the model constructed to account for the incoming particles to the monolayer as presented in the main manuscript. The dashed black line shows $N\sim\tilde{t}^{2.5}$, since it is expected from simulating deposition models that $\bar{h}$ and $N$ scale in the same way with time.}  \label{fig: ht_Nt} 
\end{figure*}

The relationship between the monolayer profile evolution and the incoming particle rate is not obvious a priori. In Fig. \ref{fig: ht_Nt} we show the typical evolution of the mean height of the monolayer $\bar{h}(t)$ (Fig. \ref{fig: ht_Nt}a). We can see that a power law with exponent 2.5 describes the growth of the monolayer for 5 mM and 10 mM for the full time range. The case of 50 mM is more complex since, as discussed in the main text, a large percentage of particles do not reach the monolayer and are found adhered to the glass slide. This effect becomes more important as the contact angle is reduced and the particle-substrate distance reduces, and therefore the monolayer growth slows down as $\bar{t}\rightarrow 1$.

The total number of particles arriving to the monolayer is shown in Fig. \ref{fig: ht_Nt}b-d. We have observed the same scaling between the monolayer growth $\bar{h}(t)$ and the total number of particles $N(t)$ in our simulations of deposition models (see details further below). Therefore, the dashed black line in the plots represents the curve relation $N\sim\tilde{t}^{2.5}$. This curve
overlaps with the experimental data for a large part of the domain, but fails to capture the initial stages. In the next sections we describe with more details the model we have constructed to account for the incoming particle rate.

\subsection{Simulations of Poissonian and Ballistic deposition models: Interface roughness}

The growth of complex interfaces was an intense topic of research in the early 80s since it was expected that simple particle deposition models could reproduce patterns observed in nature. These deposition models were based on particles arriving randomly to an interface where they will adhere irreversibly following certain rules. The growth of such interfaces was studied intensely but special interest was devoted to study the roughness of the formed interface for its potential reactive properties. In order to verify if the growth of our particle monolayer follows any of these deposition models, we have simulated such processes using the particle rate found in our experiments ($N \sim t^{2.5}$, see Fig. \ref{fig: ht_Nt}a) and computed the scaling of the roughness $w$ versus the mean height of the monolayer $\bar{h}$. In both models we will employ $L=$ 5000 columns that will be filled in each iteration $t=[1,2,...,t_{max}]$ by $N(t)=0.1\times t^{2.5}$ particles distributed randomly. Each data point shown is the averaged value of $w$ over 100 simulations ran under the same conditions.

\begin{figure*}
  \includegraphics[width=1\textwidth,trim={5 10 5 10}]{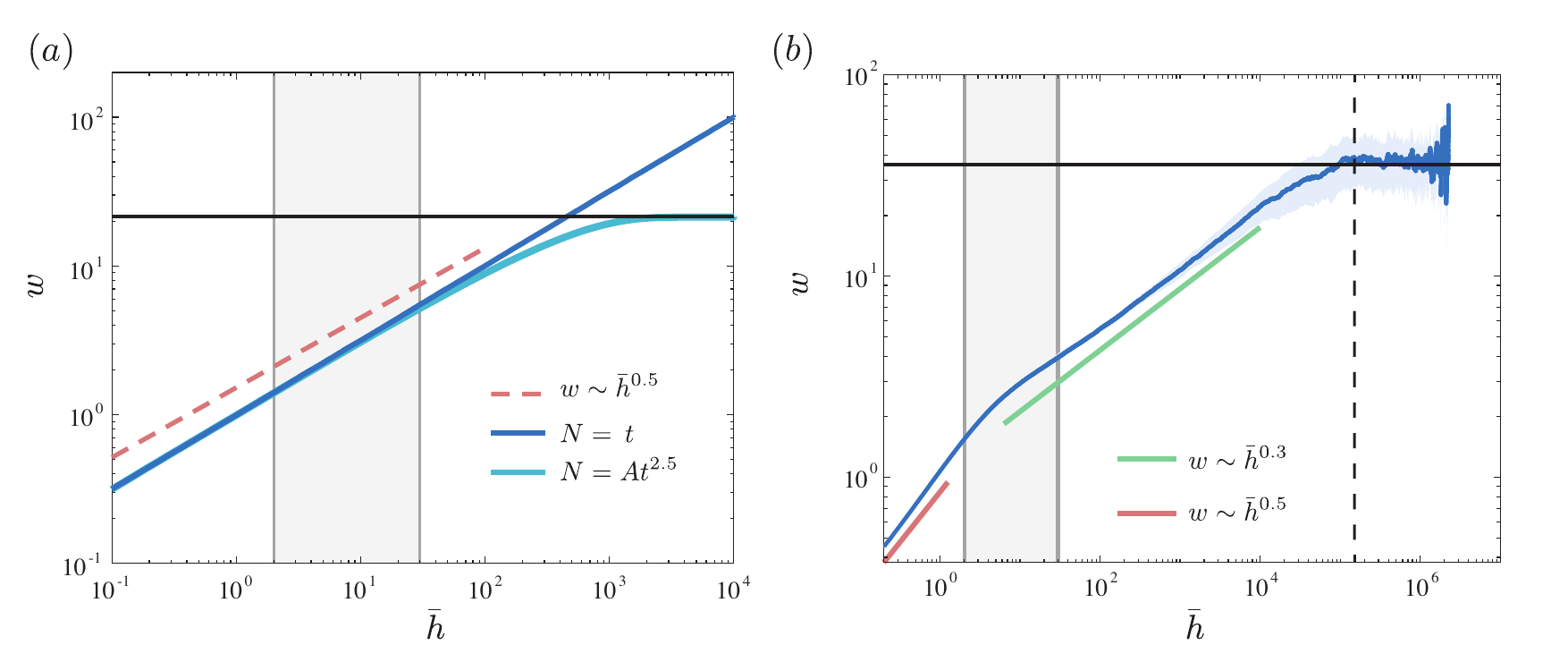}
\vspace{-5mm} \caption{The interface roughness $w$ as a function of the mean height $\bar{h}$, obtained from 100 numerical simulations of the classical deposition models: (a) random or Poissonian deposition and (b) Ballistic deposition. Both are 1D systems with size $L=5000$. The vertical shaded area represents the region of $\bar{h}$ captured experimentally. In (a) simulations are performed for a constant particle rate ($N = t$, to validate the algorithm) and for increasing particle influx as in the experiments ($N=0.1\times t^{2.5}$). (b) The Ballistic deposition model shows larger variability in the larger values of $\bar{h}$ which is represented by the shaded region around the main curve showing the standard deviation of the 100 simulations. Interestingly, despite the variable particle rate ($N=0.1\times t^{2.5}$), the roughness follows the same scaling found by Family and Vicsek \cite{Family1986scaling}.}  \label{fig: Sim_roughness} 
\end{figure*}

In the Poissonian or random deposition model  \textit{``particles simply fall until they reach the top of the column in which they were dropped or they reach the substrate. At this point they stop and become part of the aggregate''} \cite{Family1990}. Figure \ref{fig: Sim_roughness}a shows $w$ vs $\bar{h}$ for the simulated random deposition cases with constant particle rate ($N = t$) and variable particle rate. The constant particle rate shows a robust scaling $w\sim \bar{h}^{0.5}$ for 5 decades (as expected from the results in the literature \cite{Family1985,Family1990}), while the results for the variable rate saturates to $w\sim 20$ for $\bar{h} \sim 10^3$. We assume this saturation in the random deposition occurs due to the large amount of particles being delivered in each iteration for large $t$. In any case, that area should not appear in our current experiments which only cover the region $2 <\bar{h} < 30$ (shaded region).

The Ballistic deposition model is one of the models following the Family-Vicsek dynamic scaling \cite{Family1985}, by which $w \sim t^n$ for moderate $t$, and it saturates to $w \sim L^m$ as $t\rightarrow L^{m/n}$. In the Ballistic deposition model ``\emph{particles rain down onto the substrate following straight-line trajectories in the columns in which they were dropped until they first encounter a particle in the deposit. This can be a particle at the top of the same column, or a particle in one of the nearest-neighbor columns. At this point the particle stops and becomes a permanent part of the deposit}'' \cite{Family1990}. The simulations of Family and Vicsek \cite{Family1985} yielded the values $n=0.3$, $m=0.42$ and interestingly, our simulations with variable incoming particle yielded identical values. In Fig. \ref{fig: Sim_roughness}b we show an continuous horizontal black line at $w = L^{0.42} = 35.77$, and vertical dashed black line at $\bar{h} = L^{0.42/0.3} = 150854$, which confirm the Family-Vicsek scaling for $w$ vs $\bar{h}$ even for non-constant particle rate. The region of $\bar{h}<1$ shows a different scaling, which is also nicely explained by Family in another paper \cite{Family1986scaling}: ``\emph{Initially, before a single layer of particles has been deposited, the diffusion process is unimportant, because it does not change the random placing of the particles on the seed particles and consequently the surface width varies as the square root of $h$}''. Despite being unimportant, we decide to plot it since our experimental range covers part of that initial square root behaviour. Note that if any of our experiments would follow a Ballistic deposition process, it would be difficult to measure it accurately since our experimental range covers part of the transition from a this initial square root process ($\bar{h}<1$) to the pre-saturation regime for $\bar{h}<L^{m/n}$. 

Another interesting observation is that, by limiting the simulations to a maximum number of particles equal to the number found in the experiments, we find a final mean height $\bar{h}_F$ of \SI{26}{\micro\meter} for the Poissonian case and \SI{51}{\micro\meter} for the Ballistic case, which are both of the same order of magnitude as obtained in the experiments for low initial salt concentration and high initial salt concentration respectively.

\subsection{Alternative data processing $w(t)$ vs $\bar{h}$}

\begin{figure*}
  \includegraphics[width=1\textwidth,trim={1 10 1 5}]{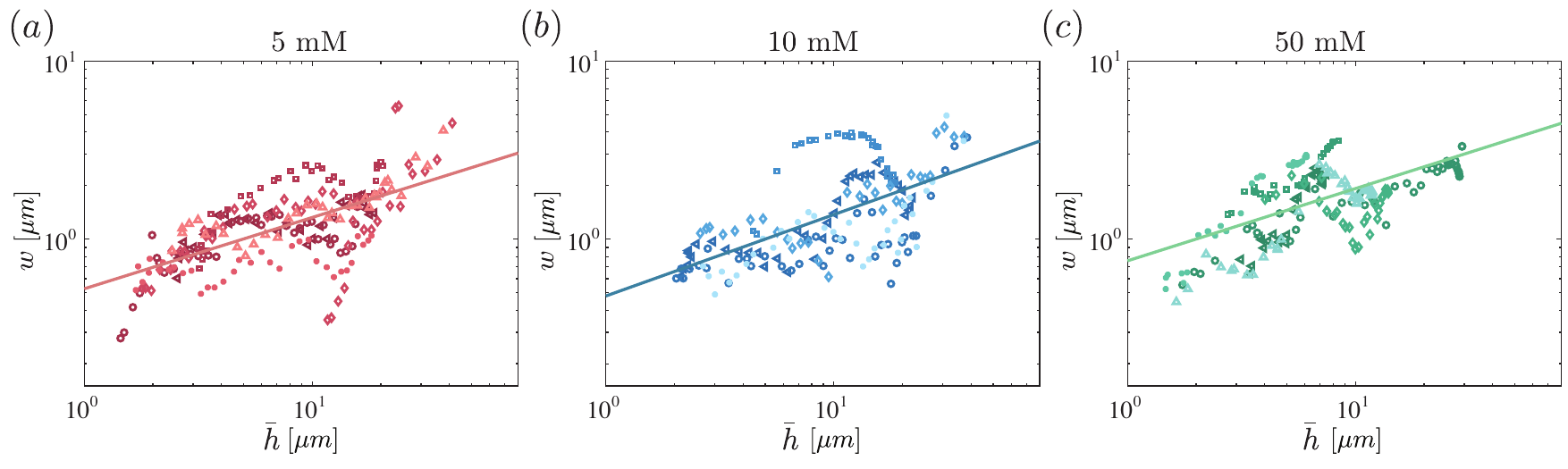}
\caption{The roughness of the monolayer contours $w$ as a function of the mean height $\bar{h}$. Figure (a) shows 5 mM, (b) 10 mM and (c) 50 mM. Different markers indicate different experiments. Curves for the best fit of $w\sim \bar{h}^n$ for each salt concentration are included in the figures: $n(\mathrm{5~mM})= 0.40\pm 0.03$, $n(\mathrm{10~mM})= 0.46\pm 0.05$ and $n(\mathrm{50~mM})= 0.41\pm0.04$. \label{fig: Roughness_nobin}}
\end{figure*}

The experimental data for the roughness has been obtained using 5 datasets per salt concentration, covering a decade in $\bar{h}$. Given the stochastic nature of the data, the strategy followed processing the data to test whether $w(\bar{h})$ follows a power-law is crucial. Here we show the results of a different fitting strategy as the one shown in the main text. Instead of binning the data and perform a linear regression to $\log w$ vs $\log \bar{h}$, we perform the linear regression directly to the \emph{raw} data. The results are shown in Fig. \ref{fig: Roughness_nobin} for the three different initial salt concentrations. The exponents obtained from the linear regression are $n(\mathrm{5~mM})= 0.40\pm 0.03$, $n(\mathrm{10~mM})= 0.46\pm 0.05$, $n(\mathrm{50~mM})= 0.41\pm0.04$. The trend obtained by this fit seemed counter-intuitive to us: 5 mM and 50 mM yield almost identical roughness exponent values in spite of their clear different behavior. By making a linear regression without any filtering, it seems that the fit has taken too much weight from those rogue data points laying in the tails of the probability distribution. However, we decided not to apply any criterion for defining outliers that would be removed from the dataset. Instead, to remove any bias from our side, we choose to perform a logarithmic binning, followed by a weighted linear regression using the number of points per bin as weights. This strategy has resulted in more coherent values that have been shown in the main text.

\subsection{Transport equation for incoming particles along the droplet surface}

To obtain a more accurate prediction of the incoming particle rate, we model the particle distribution at the interface of the droplet by solving numerically a simplified interfacial transport equation for the particle concentration $C(s,t)$. The mechanism proposed is the simplest possible: particles are captured by the droplet surface as the contact angle decreases (source term $J(s,t)$) and advected along the surface due to the Marangoni flow at an unknown velocity $u(s,t)$ until they arrive at the monolayer, located at the contact line. This transport equation reads

\begin{equation}
    \label{dC_dt}
\frac{\partial C}{\partial t} = J(s,t)-\nabla\cdot (uC) =  J(s,t)-u\frac{\partial C}{\partial s} -C\frac{\partial u}{\partial s}. 
\end{equation}

\begin{figure}
  \includegraphics[width=0.5\textwidth]{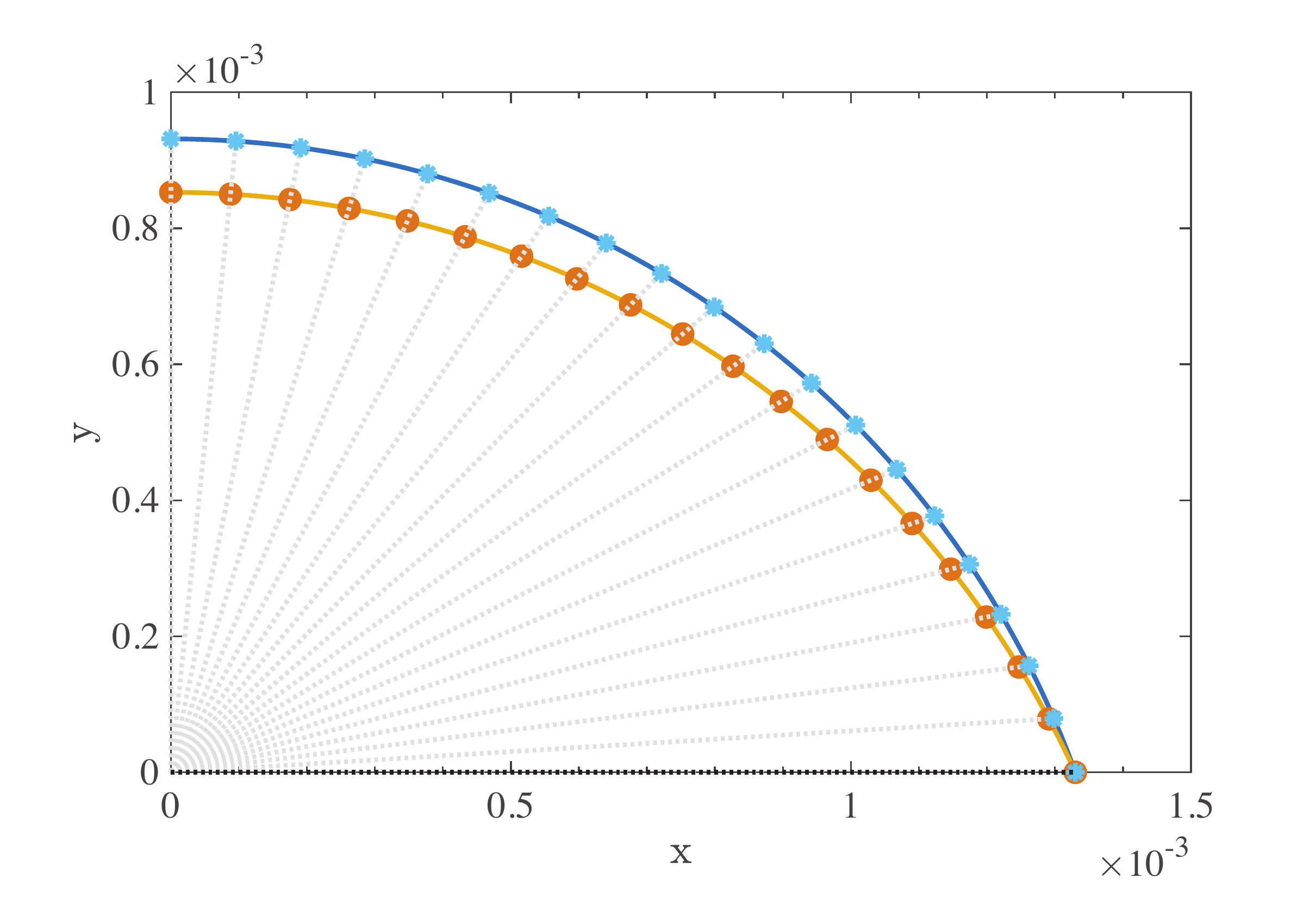}
\vspace{-3mm} \caption{A schematic representation is given to show how the change in local volume is implemented into the numerical code. In this schematic the number of grid points is 20. The two spherical caps represent $t=$ 0 s  (blue) and $t$ is 100s (orange), with respectively $\theta$ = 70 and 65 degrees. The grid points at $t$=100 are equally spaced along the arc, and the corresponding points at the spherical cap of $t$=0 are shown.  }  \label{fig: ChangeH} 
\end{figure}

Given the lack of experimental data or theoretical/numerical solutions on $u(s,t)$, we decide to use a constant value based on our own observations and comparable to those found by Marin \emph{et al.} \cite{Marin2019}. The amount of particles trapped by the droplet surface per unit time and area is represented by $J(s,t)$, which acts as a source term in this equation. Equation \ref{dC_dt} is solved in 1D along the interface (for $0\leq s \leq L$) as the system is axisymmetric around the apex of the droplet. 
The assumptions which are made in order to solve this equation are (1) an initial homogeneous particle distribution in the bulk of the droplet, represented by the bulk concentration $C_B$. (2) The particles are captured at the surface due to the surface area decrease during the evaporation while keeping a constant base radius. Under such conditions, $J(s,t)$ can be obtained directly using the model developed by Popov \cite{ Popov2005}: we solve the differential equation for $\theta(t)$ for our experimental conditions: $R_i =$ 1.3 mm, $\theta_i$ = 70$^{\circ}$ and $c_s$ = 2.08$\cdot 10^{-2}$ kg/m$^3$\cite{Gelderblom2011}. Solving this equation for $\theta$ using a forward Euler method, and combining these two assumptions we can write the change of particles captured at the interface as

\begin{equation}
    \label{dN_dt}
    \frac{dN_s}{dt} = -C_{B}\frac{dV}{dt} ,  
\end{equation}

with $C_B = 1.5\cdot 10^6$ (\# part/m$^3$), using our initial particle concentration of 0.01 $\%$w. To implement the spatial dependence of the influx of particles we discretize the system into smaller volumes. A schematic of this discretization is shown in Fig. \ref{fig: ChangeH}. As locally the volume of the evaporated compartment changes along $s$, a different number of particles will be swept by the moving interface. We use equation \ref{dN_dt} to convert this local volume loss into a number of particles. When we also take the area of the local segment into account, we can calculate the local particle flux $J(s)$ for each time step. The value of $J$ decreases for increasing $s$, which means it is the largest near the apex of the droplet. This is to be expected, since the apex is the only point in the droplet surface following a direct downward motion (same direction as gravity) and consequently it has the highest downward velocity.

When we implement all geometrical equations to describe the spherical cap shape at every time step, combined with the fact that our numerical domain shrinks in time ($L$ is a function of time, with $L(t_F)= R_i$), we can solve Eq. \ref{dC_dt} and obtain the particle concentration $C(s,t)$. The boundary conditions which are used to solve this equation are $C(0,t)$ = 0 and and an outflux (Neumann) condition at $C(L,t)$. From the solution of the concentration field we can extract a number of particles which is located at the contact line, i.e. at the end of the domain ($s = L$). We also take into account that our experimental measurement section is only \SI{25}{\micro\meter} wide by introducing a correction $\phi = 2\pi R_i/$\SI{25}{\micro\meter}, in order to compare this numerical solution to the experimentally obtained values. 

\bibliographystyle{apsrev4-1}
\bibliography{PhD-Salt3}

\end{document}